# Identification of Constitutive Parameters Governing the Hyperelastic Response of Rubber by Using Full-field Measurement and the Virtual Fields Method

A. Tayeb[1,2], J.-B. Le Cam[1,2], M. Grédiac[3], E. Toussaint[3], F. Canévet[4], E. Robin[1,2], X. Balandraud[3]

1. Univ Rennes, CNRS, IPR (Institut de Physique de Rennes) - UMR 6251, F-35000 Rennes, France
2. LC-DRIME, Joint Research Laboratory, Cooper Standard - Institut de Physique UMR 6251, Campus de Beaulieu, Bât. 10B, 35042 Rennes Cedex, France.
3. Université Clermont Auvergne, CNRS, Sigma Clermont, Institut Pascal, F-63000 Clermont-Ferrand, France
4. Cooper Standard France, 194 route de Lorient, 35043 Rennes - France.

## ABSTRACT

In this study, the Virtual Fields Method (VFM) is applied to identify constitutive parameters of hyperelastic models from a heterogeneous test. Digital image correlation (DIC) was used to estimate the displacement and strain fields required by the identification procedure. Two different hyperelastic models were considered: the Mooney model and the Ogden model. Applying the VFM to the Mooney model leads to a linear system that involves the hyperelastic parameters thanks to the linearity of the stress with respect to these parameters. In the case of the Ogden model, the stress is a nonlinear function of the hyperelastic parameters and a suitable procedure should be used to determine virtual fields leading to the best identification. This complicates the identification and affects its robustness. This is the reason why the sensitivity-based virtual field approach recently proposed in case of anisotropic plasticity by Marek et al. (2017) [1] has been successfully implemented to be applied in case of hyperelasticity. Results obtained clearly highlight the benefits of such an inverse identification approach in case of non-linear systems.

**Keywords:** Inverse identification, virtual fields method, sensitivity-based virtual fields, hyperelasticity, Digital image correlation

## Introduction

The constitutive parameters of hyperelastic models are generally identified from several homogeneous tests, typically uniaxial tension (UT), pure shear (PS) and equibiaxial tension (EQT). An alternative methodology consists in performing only one heterogeneous test [2] [3] [4] [5] that induces a large number of mechanical states at the specimen's surface. The resulting heterogeneous strain fields are generally measured by the Digital Image Correlation (DIC) technique during the loading. Among the different identification methodologies available, the Virtual Field Method (VFM) has been successfully applied to hyperelasticity in [3]. In this work, linear systems are obtained due to the linearity of the stress with respect to the constitutive parameters (see the models in [6] and [7]). When the system becomes non-linear, typically for the Odgen model [8], a statistical analysis can be carried out for optimizing the choice of the virtual fields. This complicates the identification procedure and affects its robustness. This is the reason why we also applied here to hyperelasticity the sensitivity-based virtual field approach recently proposed in case of anisotropic plasticity by [1].

## Theoretical background

Assuming a plane stress state and large strains, the principle of virtual work can be expressed as follows in the Lagrangian configuration:

$$-e_0 \int_{S_0} \Pi(X,t) : \frac{\partial U^*}{\partial X}(X,t) dS_0 + e_0 \int_{\partial S_0} (\Pi \cdot N) \cdot U^*(X,t) dL_0 = 0, \tag{1}$$

where $\Pi$ is the first Piola-Kirchhoff stress tensor, $e_0$ is thickness of the solid, $S_0$ is the surface of the solid in the normal direction to the thin dimension and $\partial S_0$ its boundary. They are measured in the reference configuration chosen here as the undeformed state. $X$ are the coordinates and $N$ denotes the normal vector to the edge.

**HYPERELASTICITY**

For hyperelastic materials, the mechanical behavior is described by the strain energy density $W$ relating the stress to the strain through the principle stretches $(\lambda_1, \lambda_2, \lambda_3)$ or the first two principal invariants of the left Cauchy-Green strain tensor ($I_1$ and $I_2$). Assuming that the material is incompressible, the first Piola-Kirchhoff stress tensor for such material reads:

$$\Pi = -pF^{-t} + \frac{\partial W}{\partial F}, \tag{2}$$

where $p$ is an indeterminate coefficient due to incompressibility, $F$ is the deformation gradient tensor and $\bullet^t$ designates the transpose of a second-order tensor. For the Mooney model [6], the strain energy density writes:

$$W = c_1(I_1 - 3) + c_2(I_2 - 3), \tag{3}$$

where $c_1$ and $c_2$ are the constitutive parameters to be identified. Combining eqs. (2) and (3) and replacing $\Pi$ by its expression in eq. (1) leads to the expression of the principle of virtual work:

$$c_1 \int_{S_0} \Theta : \frac{\partial U^*}{\partial X} dS_0 + c_2 \int_{S_0} \Lambda : \frac{\partial U^*}{\partial X} dS_0 = \int_{\partial S_0} (\Pi \cdot N) \cdot U^* dL_0, \tag{4}$$

where $\Theta$ and $\Lambda$ are two functions of the principle stretches. Using eq. (4) with two independent displacement virtual fields leads to the following linear system

$$\mathbf{Ac} = \mathbf{B}$$
with
$$\mathbf{A} : \begin{bmatrix} \int_{S_0} \Theta : \frac{\partial \mathbf{U}^{*(1)}}{\partial \mathbf{X}} dS_0 & \int_{S_0} \Lambda : \frac{\partial \mathbf{U}^{*(1)}}{\partial \mathbf{X}} dS_0 \\ \int_{S_0} \Theta : \frac{\partial \mathbf{U}^{*(2)}}{\partial \mathbf{X}} dS_0 & \int_{S_0} \Lambda : \frac{\partial \mathbf{U}^{*(2)}}{\partial \mathbf{X}} dS_0 \end{bmatrix} \tag{5}$$
and
$$\mathbf{c} = \begin{Bmatrix} c_1 \\ c_2 \end{Bmatrix}, \mathbf{B} = \begin{Bmatrix} \int_{\partial S_0} (\mathbf{\Pi} \cdot \mathbf{N}) \cdot \mathbf{U}^{*(1)} dL_0 \\ \int_{\partial S_0} (\mathbf{\Pi} \cdot \mathbf{N}) \cdot \mathbf{U}^{*(2)} dL_0 \end{Bmatrix}$$

After inversion, this system gives the two constitutive parameters $c_1$ and $c_2$. The second model considered in the present study is due to [8]. In this case, the strain energy density reads

$$W = \sum_{i=1}^{N} \frac{2\mu_i}{\alpha_i^2}\left(\lambda_1^{\alpha_i} + \lambda_2^{\alpha_i} + \lambda_3^{\alpha_i} - 3\right), \tag{6}$$

where $\mu_i, \alpha_i; i = 1..N$ are the constitutive parameters. From eqs. (2) and (6), the eigenvalues of the Piola-Kirchhoff stress tensor are defined by

$$\Pi_i = \frac{\partial W}{\partial \lambda_i} - \lambda_i^{-1} p. \tag{7}$$

The principle of virtual work of eq. (1) becomes in this case

$$-\int_{S_0} \left(\Pi_1 \cdot U_{u,u}^* + \Pi_2 \cdot U_{v,v}^*\right) dS_0 + \int_{\partial S_0} (\mathbf{\Pi} \cdot \mathbf{N}) \cdot \mathbf{U}^*(\mathbf{X},t) dL_0 = 0, \tag{8}$$

where $(u,v)$ is the principle basis of the strain tensor. In this basis, the cost function to be minimized to find the constitutive parameters can be written as follows

$$\mathbf{f}(\chi) = \sum_{j=1}^{nVF} \left[ \sum_{t=1}^{nTime} \left( \sum_{i=1}^{nPts} \left(\Pi_1(\chi) \cdot U_{u,u}^{*i(j)} + \Pi_2(\chi) \cdot U_{v,v}^{*i(j)}\right) S^i - W_{ext}^* \right)^2 \right] \tag{10}$$

$nVF$, $nTime$ and $nPts$ denote respectively the number of independent virtual fields, the time steps and the number of Zones of Interest (ZOIs).

In this paper, only the first order $(N = 1)$ of this model is considered. The identification of the constitutive parameters is carried out by minimizing the cost function $\mathbf{f}$.

**CHOICE OF THE VIRTUAL DISPLACEMENT FIELDS**

Since an infinite number of kinematically admissible virtual fields $\mathbf{U}^*$ satisfies the principle of virtual work in Eq. (1), the choice of a set of independent virtual fields remains a typical issue. The case of linear elasticity is discussed in [9]. We address here the most challenging case of hyperelasticity.

RANDOM VIRTUAL DISPLACEMENT FIELDS

To deal with hyperelastic materials, different approaches in the generation of independent virtual displacement fields should be applied. To the authors' knowledge, the virtual fields method was first applied to hyperelastic materials in [3]. Motivated by a noise-sensitivity study, a set of random virtual displacement fields was generated. The procedure relies on the division of the region of interest in the sample into 12 quadrangular sub-domains over which piecewise virtual fields are defined. Random values for the displacement at the nodes are then generated and the set of virtual fields leading to the best identification is chosen. The displacement is approximated in each sub-domain by using four-noded quadrangular finite elements formulations [10]. In [3], these random virtual displacement fields were used for two hyperelastic models for which the application of the virtual fields method led to a linear system. In this case, to ensure the independence of the virtual fields generated, the criterion was a good conditioning of the system defined in Eq. (5). However, for models for which the virtual field method does not lead to a linear system, such as the Ogden model, a statistical study should have been done to generate independent virtual fields, which makes identification more complicated and less robust and thus requires the development of alternative strategies.

SENSITIVITY-BASED VIRTUAL DISPLACEMENT FIELDS

In a recent work [1], a new procedure for generating independent virtual displacement fields was employed for the identification of the constitutive parameters in the case of an anisotropic plastic material in the small strain domain. The method is based on the sensitivity of the stress to small changes of the constitutive parameters. The virtual displacement fields are then generated proportionally to the stress sensitivity fields through a finite element-like approach. The method was then extended to finite strain for anisotropic plasticity in [11]. For finite strains, the stress sensitivity field is defined by

$$\delta \mathbf{\Pi}^{(i)}(\chi, t) = \mathbf{\Pi}(\chi + \delta\chi_i, t) - \mathbf{\Pi}(\chi, t), \quad (11)$$

where $0.1\chi_i \leq \delta\chi_i \leq 0.2\chi_i$ is the sensitivity of the $i^{th}$ parameter, which numerical values for the parameter picked in the literature. Note that the stress sensitivity field in Eq. (11) gives the influence of each constitutive parameter in the global response of the material at each point since the experiment used is heterogeneous. Therefore, the virtual displacement fields were generated proportionally by setting the stress sensitivity fields with the following expression:

$$\delta \mathbf{\Pi}^{(i)}(\chi, t) = \mathbf{B}_{glob} \mathbf{U}^{*(i)}, \quad (12)$$

where $\mathbf{B}_{glob}$ is the global strain-displacement matrix from a virtual mesh generated *a priori*. This matrix is obtained by assembling of the elementary strain-displacement matrix obtained directly from the derivation of the shape functions with respect to the coordinates. $\mathbf{U}^{*(i)}$ in eq. (12) designates the virtual displacement field corresponding to the $i^{th}$ constitutive parameter. Note that this virtual displacement field is a *test function* and has no physical meaning. In practice, matrix $\mathbf{B}_{glob}$ should be modified to account for the boundary conditions of the region of interest (ROI). Typically, for edges where external loading is unknown, a null displacement should be imposed. Therefore, a new matrix $\bar{\mathbf{B}}_{glob}$ is obtained from the original matrix $\mathbf{B}_{glob}$. The virtual displacement field is then given by

$$\mathbf{U}^{*(i)} = pinv(\bar{\mathbf{B}}_{glob}) \delta \mathbf{\Pi}^{(i)}(\chi, t), \quad (13)$$

where $pinv$ designates the pseudo inverse operator. Once the virtual displacement field is obtained, its gradient needed in the principle of virtual work is computed using the classic equation obtained with the finite elements method

$$\frac{\partial \mathbf{U}^{*(i)}}{\partial \mathbf{X}} = \mathbf{B}_{glob} \mathbf{U}^{*(i)}. \quad (14)$$

The contribution of each constitutive parameter to the response of the material is very different and unique. Therefore, a scaling in the cost function should be added (see [1] and [11]).

**EXPERIMENTS**

The material used in this study is a carbon black filled natural rubber. The sample is shown in **Erreur ! Source du renvoi introuvable.**. It is 105 mm in length and 2 mm in thickness. The experimental setup is presented in Figure 2**Erreur ! Source du renvoi introuvable.**. It is composed by a home-made biaxial testing machine and a digital camera. The four independent actuators were linked to have the same movement such that the specimen center was motionless during the test. Hence, a reference point is obtained in the center of the sample with respect to the correlation procedure. A displacement of $70\,mm$ was applied to each branch at a loading rate of $150\,mm/min$ which corresponds to a value of $\lambda_{max}$ of around $3.4$. During the mechanical test, images of the specimen surface were stored at a frequency of $5\,Hz$ using an IDS camera equipped with a $55\,mm$ telecentric objective. The charge-coupled device (CCD) sensor of the camera has $1920 \times 1200$ joined pixels. The displacement field at the surface of the specimen was determined using the digital image correlation (DIC) technique. The correlation process is achieved thanks to the SeptD software [12]. The spatial resolution, defined as the smallest distance between two independent points, was equal to $10\,pixels$. A rectangular region on one branch of the specimen and including the specimen centre is sufficient to apply the identification procedure described above. The rectangular region of interest (*R.O.I.*) is represented in **Erreur ! Source du renvoi introuvable.**.

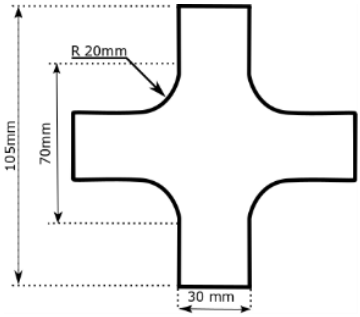
Figure 1: Sample geometry

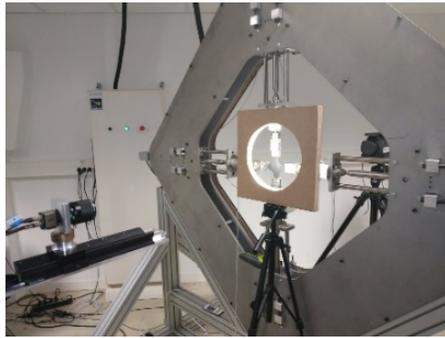
Figure 2: Experimental setup

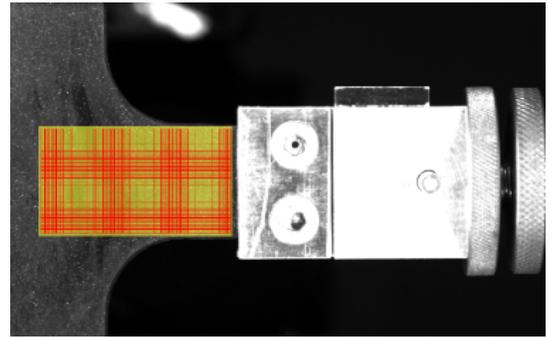
Figure 3: ROI with ZOIs of 10 by 10 px

**RESULTS**

EXPERIMENTAL KINEMATIC FIELDS

The displacement field obtained from the SeptD software is smoothed using a mean centered filter. The values in the zones of interest (*ZOIs*) where the correlation could not be achieved were interpolated. The displacement gradient fields are presented in Figure 4 for the rectangular *ROI*. These data were smoothed using the same filter. The data obtained experimentally were used in the identification of the constitutive parameters for random and sensitivity-based virtual displacement fields.

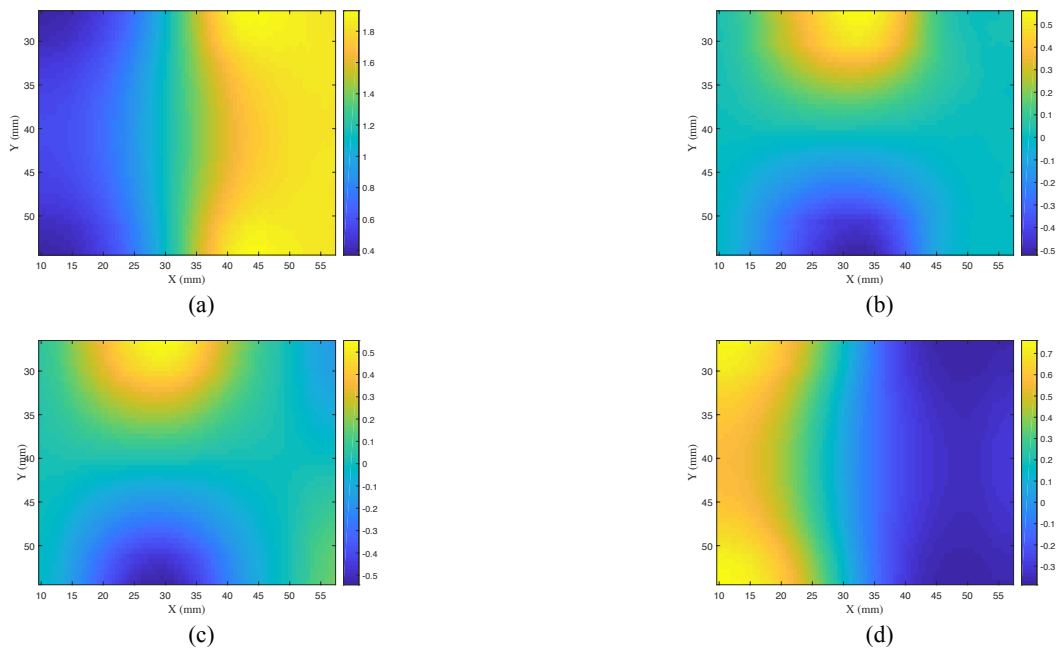
Figure 4: Experimental displacement gradient fields

RESULTS

For the randomly generated virtual fields, the virtual displacement fields for the Mooney model were chosen in such a way that the conditioning of the matrix $\mathbf{A}$ of eq. (5) is greater than $0.3$. For the Ogden model, no criterion is available for the choice of the virtual displacement fields. Hence, a wide number of virtual fields were generated and used in the identification procedure. The parameters identified using this approach are reported in Table 1.

| Model | Parameters | Value |
| --- | --- | --- |
| Mooney [6] | $c_1$ | $0.229$ MPa |
| - | $c_2$ | $9.4 \, 10^{-4}$ MPa |
| Ogden [8] | $\mu_1$ | $0.68$ MPa |
| - | $\alpha_1$ | $1.4$ |

Table 1: Identified hyperelastic constitutive parameters using random virtual displacement fields

For the sensitivity based virtual fields, first, a virtual mesh is generated. It can be different from the correlation grid. Then, the virtual fields are generated proportionally to the stress sensitivity fields. The values of the parameters identified are reported in Table 2. Unlike for random virtual fields, the parameters for all the models considered are in the range of those reported in the literature. Note that the parameters of Table 2 are obtained for several simulations with different sensitivity parameters, i.e., for different virtual fields. Furthermore, the mean values for the parameters did not affect the final result of the identification. In fact, the mean values for each parameter could change within a given range without affecting the final result of the identification. For the Ogden model, the least square error (the value of the objective function at the end of the identification) is about $1.5 \, 10^{-5}$.

| Model | Parameters | Value |
| --- | --- | --- |
| Mooney [6] | $c_1$ | $0.22$ MPa |
| - | $c_2$ | $1.9 \, 10^{-2}$ MPa |
| Ogden [8] | $\mu_1$ | $0.46$ MPa |
| - | $\alpha_1$ | $2.11$ |

Table 2: Parameters identified using sensitivity based virtual fields

To evaluate the accuracy of the parameters identified, the biaxial experiment used in this work was simulated using Abaqus software for a plane stress problem with the parameters of Table 1 and Table 2. For each set of parameters, the resulting force predicted in every branch of the sample was compared to the experimental force measured. The results of the comparison are shown in Figure 5 in which *SBVF* and *RVF* refer to sensitivity-based virtual fields and random virtual fields, respectively. For the *RVF* method, the Mooney model appears to have a good result for a maximum stretch up to $2.7$, which is the usual range for this model. However, the Ogden model overestimates the force in the branch for the whole experiment. This is due to the choice of the virtual displacement fields, which was done randomly and no criterion was found in its selection for this model. For the *SBVF* method, the Mooney model has a good prediction for the experimental force for a principal stretch up to $2.6$. The Ogden model has a better result for wider strain range corresponding to a principle stretch up to $3$. The results of these two models are very satisfactory given that they do not take into account the stress-hardening phenomenon. Hence, the capacity of the *SBVF* method in the generation of the virtual displacement fields is illustrated here in the case of hyperelastic behavior.

**CONCLUSIONS**

In this study, the Virtual Fields Method (VFM) was applied to identify constitutive parameters of hyperelastic models from a heterogeneous test in the cases of linear and non-linear relationships between the stress and the constitutive parameters to be identified. In the former case, the Mooney model was considered and the virtual field were randomly generated. For the latter case, the Ogden model was used and a sensitivity-based virtual fields approach inspired from a recent work due to [1] for anisotropic plasticity was applied to choose the virtual fields. Results obtained with the two approaches clearly highlight the benefits of using the sensitivity based virtual fields approach for identifying the constitutive parameters in case of non-linear systems.

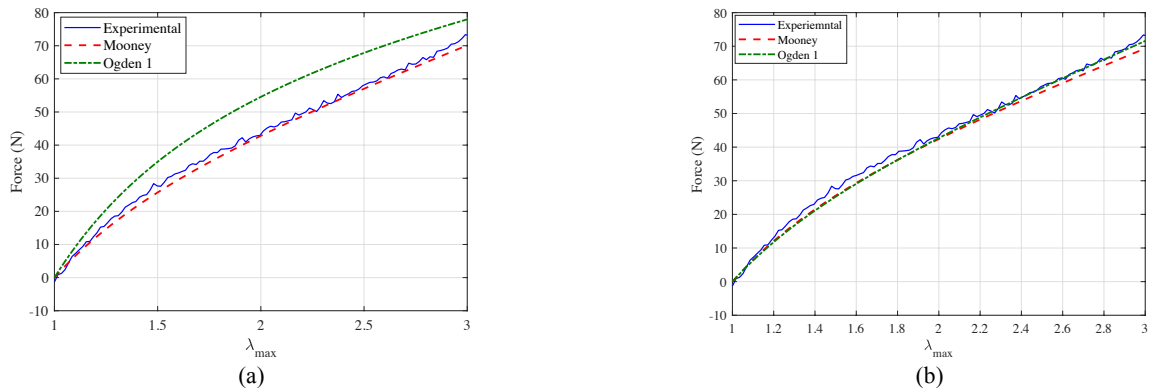

Figure 5: Force obtained from finite element simulations compared to experimental force